\documentclass[twocolumn,numberedappendix,twocolappendix,appendixfloats]{openjournal_dhayaa}

\usepackage{comment}
\usepackage{amsmath}	

\usepackage{graphicx}	
\graphicspath{{./}{./plots/}}

\usepackage{xcolor}
\usepackage{xspace}
\usepackage{amsmath}
\usepackage{enumitem}

\usepackage{natbib}
\setcitestyle{aysep={}}

\definecolor{xlinkcolor}{cmyk}{1,1,0,0}

\usepackage[
    colorlinks=true,
    urlcolor=xlinkcolor,
    anchorcolor=xlinkcolor,
    citecolor=xlinkcolor,
    filecolor=xlinkcolor,
    linkcolor=xlinkcolor,
    menucolor=xlinkcolor,
    linktocpage=true,
    pdfproducer=medialab,
    pdfa=true,
]{hyperref}

\usepackage{fontawesome5}
\usepackage{textpos}
\usepackage{lineno}

\definecolor{orcidlogocol}{HTML}{A6CE39}
\definecolor{purple}{RGB}{128, 0, 128}
\definecolor{maroon}{RGB}{128, 0, 0}

\newcommand{\OrcidID}[1]{ \href[urlcolor = red]{https://orcid.org/#1}{\textcolor{lightgray}{\faOrcid}}}
\newcommand{\OrcidIDName}[2]{\href{https://orcid.org/#1}{#2}}

\newcommand{\nobs}{36\xspace}
\newcommand{\ntotideal}{\ensuremath{89 \pm 20}\xspace}
\newcommand{\nideal}{\ensuremath{53}\xspace}
\newcommand{\ntotmeas}{\ensuremath{83 \pm 18}\xspace}
\newcommand{\nmeas}{\ensuremath{47}\xspace}
\newcommand{\ntotcorr}{\ensuremath{67 \pm 14}\xspace}
\newcommand{\ncorr}{\ensuremath{31}\xspace}

\newcommand{\software}[1]{\textsc{#1}\xspace}
\newcommand{\simple}{\software{simple}}

\begin{document}


\title{Predictions for the Detectability of Milky Way Satellite Galaxies and Outer-Halo Star Clusters with the Vera C.\ Rubin Observatory}
\shortauthors{Tsiane, Mau, Drlica-Wagner et al.}
\shorttitle{Detectability of Milky Way Satellites by Rubin Observatory}

\author{\OrcidIDName{0009-0008-6557-2065}{Kabelo~Tsiane\,$^\dagger$}}

\affiliation{Department of Physics, University of Michigan, Ann Arbor, MI 48109, USA}
\affiliation{Department of Astronomy and Astrophysics, University of Chicago, Chicago, IL 60637, USA}
\author{\OrcidIDName{0000-0003-3519-4004}{Sidney~Mau\,$^\dagger$}}
\affiliation{Department of Physics, Stanford University, 382 Via Pueblo Mall, Stanford, CA 94305, USA}
\affiliation{Kavli Institute for Particle Astrophysics \& Cosmology, P.O.\ Box 2450, Stanford University, Stanford, CA 94305, USA}
\author{\OrcidIDName{0000-0001-8251-933X}{Alex~Drlica-Wagner\,$^\dagger$}}
\affiliation{Fermi National Accelerator Laboratory, P.O.\ Box 500, Batavia, IL 60510, USA}
\affiliation{Department of Astronomy and Astrophysics, University of Chicago, Chicago, IL 60637, USA}
\affiliation{Kavli Institute for Cosmological Physics, University of Chicago, Chicago, IL 60637, USA}
\affiliation{NSF-Simons AI Institute for the Sky (SkAI), 172 E.\ Chestnut St., Chicago, IL 60611, USA}
\author{\OrcidIDName{0000-0002-3936-9628}{Jeffrey~L.~Carlin}}
\affiliation{Rubin Observatory/AURA, 950 North Cherry Avenue, Tucson, AZ, 85719, USA}
\author{\OrcidIDName{0000-0001-6957-1627}{Peter~S.~Ferguson}}
\affiliation{DiRAC Institute, Department of Astronomy, University of Washington, 3910 15th Ave NE, Seattle, WA, 98195, USA}
\author{\OrcidIDName{0000-0001-8156-0429}{Keith~Bechtol}}
\affiliation{Physics Department, 2320 Chamberlin Hall, University of Wisconsin-Madison, 1150 University Avenue Madison, WI  53706-1390}
\author{\OrcidIDName{0000-0002-1182-3825}{Ethan~O.~Nadler}}
\affiliation{Department of Astronomy \& Astrophysics, University of California, San Diego, La Jolla, CA 92093, USA}
\author{\OrcidIDName{0000-0002-8040-6785}{Annika~H.~G.~Peter}}
\affiliation{The Ohio State University, Department of Physics, Columbus, OH 43210, USA}
\affiliation{The Ohio State University, Department of Astronomy, Columbus, OH 43210, USA}
\affiliation{Center for Cosmology and Astroparticle Physics, 191 West Woodruff Avenue, Columbus, OH 43210, USA}
\author{\OrcidIDName{0000-0002-1200-0820}{Yao-Yuan Mao}}
\affiliation{Department of Physics and Astronomy, University of Utah, Salt Lake City, UT 84112, USA}
\author{\OrcidIDName{0000-0001-9342-6032}{Adam~J.~Thornton}}
\affiliation{Vera C.\ Rubin Observatory/NOIRLab, 950 N.\ Cherry Ave., Tucson, AZ 85719, USA}
\author{The LSST Dark Energy Science Collaboration}


\email{$^\dagger$ kabelo@umich.edu, smau@stanford.edu, kadrlica@fnal.gov}

\begin{abstract}
We predict the sensitivity of the Vera C.\ Rubin Observatory Legacy Survey of Space and Time (LSST) to faint, resolved Milky Way satellite galaxies and outer-halo star clusters. 
We characterize the expected sensitivity using simulated LSST data from the LSST Dark Energy Science Collaboration (DESC) Data Challenge 2 (DC2) accessed and analyzed with the Rubin Science Platform as part of the Rubin Early Science Program. 
We simulate resolved stellar populations of Milky Way satellite galaxies and outer-halo star clusters over a wide range of sizes, luminosities, and heliocentric distances, which are broadly consistent with expectations for the Milky Way satellite system.
We inject simulated stars into the DC2 catalog with realistic photometric uncertainties and star/galaxy separation derived from the DC2 data itself.
We assess the probability that each simulated system would be detected by LSST using a conventional isochrone matched-filter technique.
We find that assuming perfect star/galaxy separation enables the detection of resolved stellar systems with $M_V = 0$\,mag and $r_{1/2} = 10$\,pc with ${\geq}\,50\%$ efficiency out to a heliocentric distance of $\sim$250\,kpc.
Similar detection efficiency is possible with a simple star/galaxy separation criterion based on measured quantities, although the false positive rate is higher due to leakage of background galaxies into the stellar sample. 
When assuming perfect star/galaxy classification and a model for the galaxy--halo connection fit to current data, we predict that \ntotideal Milky Way satellite galaxies will be detectable with a simple matched-filter algorithm applied to the LSST wide-fast-deep data set.  Different assumptions about the performance of star/galaxy classification efficiency can decrease this estimate by $\sim$7\%--25\%, which emphasizes the importance of high-quality star/galaxy separation for studies of the Milky Way satellite population with LSST.
\end{abstract}


\maketitle

\section{Introduction} 
\label{sec:intro}

In the standard cosmological model comprised of cold dark matter (CDM) and a cosmological constant dark energy ($\Lambda$), galaxies form hierarchically through the assembly and mergers of dark matter halos \citep[e.g.,][]{Cole:2000}.
Ultra-faint dwarf galaxies occupy the lowest mass dark matter halos known to host stars \citep[e.g.,][]{Simon:2019, Nadler:2020}.
As the smallest and least luminous galaxies, ultra-faint dwarfs provide a unique avenue to probe the physics of galaxy formation \citep[e.g.,][]{Applebaum:2021, Munshi:2021, Kravtsov:2022}, reionization \citep[e.g.,][]{Benson:2002, Lunnan:2012, Boylan-Kolchin:2015, Graus:2019, Manwadkar:2022}, the formation of the heavy elements \citep[e.g.,][]{Ji:2016,Frebel:2023}, and the fundamental properties of dark matter \citep[e.g.,][]{Ackermann:2015, Bullock:2017, Buckley:2018,  Nadler:2021, Newton:2021, Dekker:2022}. 
While the distinction between dark-matter-dominated dwarf galaxies and baryon-dominated halo star clusters was once clear, recent observational advances have revealed new populations of ultra-faint systems of uncertain origin \citep[e.g.,][and references therein]{Pace:2024}.
Each newly discovered system increases our understanding of the Milky Way satellite population, as well as providing opportunities for unique, fortuitous discoveries among the most extreme stellar systems.

Milky Way satellite dwarf galaxies and halo star clusters (which we collectively refer to as Milky Way ``satellites'') are detected as arcminute-scale statistical over-densities of individually resolved stars \citep[e.g.,][]{Willman_2011}. 
More than 65 confirmed and candidate dark-matter-dominated satellite galaxies have been detected around the Milky Way to date \citep[e.g.,][and references therein]{Pace:2024}. When correcting for observational completeness and accounting for theoretical uncertainties, the known population of satellite galaxies is found to agree with predictions from the CDM model \citep[e.g.,][]{Kim:2018,Nadler:2020,Manwadkar:2022}. 
However, observational and theoretical arguments suggest that the current census of Milky Way satellites is incomplete, and recent models predict that the total Milky Way satellite galaxy populations consists of $\sim$100--300 systems \citep[e.g.,][]{Tollerud_2008, Koposov_2008, Walsh_2009, Jethwa:2018, Newton:2018, Kim:2018, Drlica-Wagner:2020, Nadler:2020, Manwadkar:2022, Nadler:2024}.
In addition, more than a dozen ultra-faint, compact stellar systems have been discovered in the outer Milky Way halo with a physical nature that is unclear \citep[e.g.,][]{Koposov_2008, 2013ApJ...767..101B, 2015ApJ...803...63K, Mau:2019, Cerny:2021, Cerny:2023b, Cerny:2023a, Smith:2024}.
These systems may be faint star clusters that formed in dwarf galaxies that were accreted and disrupted \citep[e.g.,][]{Malhan:2022}, or they may be an extension of the dwarf galaxy population into the small, hyper-faint regime \citep[e.g.,][]{Manwadkar:2022, Errani:2023}.

Thus far, the detection of ultra-faint Milky Way satellites has been limited by the sensitivity and sky coverage of observational surveys \citep[e.g.,][]{Koposov_2008, Walsh_2009, Drlica-Wagner:2020, Homma:2023}. The upcoming Legacy Survey of Space and Time (LSST) that will be performed by the NSF-DOE Vera C.\ Rubin Observatory is expected to greatly advance the frontier of ultra-faint satellite discovery.  LSST is an optical/near-infrared survey that is expected to reach a coadded imaging depth of $r \sim 27.5$\,mag (${\rm S/N} = 5$, point-like sources) by the end of its 10-year mission \citep{Ivezic:2019}. This sensitivity is expected to enable the detection of resolved stellar systems around the Milky Way with surface brightnesses as faint as $\mu_V \sim 32$\,mag\,arcsec$^{-2}$ over a sky area covering $\sim$20,000 deg$^2$ \citep{Tollerud_2008, Hargis_2014}. Furthermore, the location of the LSST footprint in the southern hemisphere provides additional sensitivity to satellites of the Magellanic Clouds \citep[e.g.,][]{Kallivayalil:2018, Patel:2020}.
LSST is expected to take a significant step toward completing the census of the faintest known galaxies ($M_V \sim 0$\,mag) out to the virial radius ($\sim$300\,kpc) of the Milky Way  \citep[e.g.,][]{Newton:2018, Nadler:2019b}.

Previous estimates for the observational sensitivity of LSST to ultra-faint Milky Way satellites have been based on high-level design specifications and/or the performance of precursor surveys \citep[e.g.,][]{Tollerud_2008, Hargis_2014, Nadler:2019b, Newton:2018, Manwadkar:2022}. However, these estimates often overlook systematic effects involved in searches for resolved stellar systems, which can have a non-trivial dependence on survey depth. In particular, distinguishing faint stars in the Milky Way halo from faint, barely resolvable background galaxies is a major observational challenge when conducting studies of faint, resolved stellar systems using ground-based optical surveys. Background galaxies greatly outnumber stars even at high Galactic latitudes in surveys with limiting magnitudes of $r \gtrsim 24$\,mag. Poor star-galaxy separation efficiency can therefore lead to a significant population of  misclassified background galaxies leaking into a stellar sample, which reduces the sensitivity and increases the contamination rate of satellite searches using resolved stars \citep[e.g.,][]{Fadely_2012, Sevilla-Noarbe:2018, Slater_2020}. According to \citet{Fadely_2012}, even in the idealized case where background galaxies with full width at half maximum (FWHM) $\gtrsim 0.2$ arcsec can be morphologically resolved, the number of {\it unresolved} galaxies will still outnumber Milky Way field stars at $r \gtrsim 23.5$. LSST will have a median seeing that is significantly larger than this idealized limit \citep[FWHM $\sim 0.7$\,arcsec;][]{Ivezic:2019}, leading to the risk that unresolved background galaxies could be a dominant contributor to ``point-source'' catalogs at even brighter magnitudes \citep[e.g.,][]{Slater_2020}. 

In this analysis, we make rigorous projections for the performance of LSST accounting for observational systematics and the efficiency of real-world search algorithms.
In particular, we apply a simple isochrone matched-filter algorithm \citep{Bechtol:2015, Drlica-Wagner:2020} to search for simulated Milky Way satellites injected into object catalogs generated from simulated LSST images processed with the LSST Science Pipelines\footnote{\url{https://pipelines.lsst.io/}} as part of the LSST Dark Energy Science Collaboration (DESC) Data Challenge 2 \citep[DC2;][]{Abolfathi_2021}.
Access to these data was provided through the Rubin Science Platform (RSP) in the context of Rubin Data Preview 0.1 (DP0.1).\footnote{\url{https://dp0-1.lsst.io/}}
We quantitatively measure the observational selection function, which incorporates realistic survey depth and star/galaxy separation as determined by the LSST Science Pipelines.  
We simulate the resolved stellar populations of $10^5$ Milky Way satellites and assess the detectability of each satellite based on its physical properties. 
We quantitatively evaluate the effect of star/galaxy separation on the search for resolved stellar systems around the Milky Way by studying several star/galaxy separation scenarios. 
The first scenario uses the star/galaxy separation criteria implemented in DC2 by the LSST Science Pipelines, which follows the procedure developed for Hyper Suprime-Cam Subaru Strategic Program \citep[HSC SSP;][]{Bosch:2018}. 
The second scenario assumes perfect star/galaxy classification and uses the true object classification from the input source catalogs.
Following \citet{Drlica-Wagner:2020}, we summarize our predicted LSST satellite sensitivity functions as both an analytic approximation and a machine-learning model for the detection probability as a function of physical size, heliocentric distance, and absolute magnitude.
Finally, we use this selection function in combination with the Milky Way satellite population model from \citet{Nadler:2020} to predict the population of Milky Way satellite galaxies that will be observed by LSST.
This work complements previous studies in the literature \citep[e.g.,][]{Tollerud_2008, Hargis_2014, Newton:2018, Nadler:2020, Manwadkar:2022} by providing a more rigorous, simulation-based estimate of the observational sensitivity expected from LSST.

This paper is organized as follows. In Section~\ref{sec:data}, we describe the DC2 simulations and our simulated Milky Way satellites. In Section~\ref{sec:analysis}, we briefly describe our matched-filter search algorithm and the analysis pipeline applied to the simulated LSST data. In Section~\ref{sec:osf}, we present the observational selection function derived from our study, and in Section~\ref{sec:lum_func} we make projections for the population of Milky Way satellite galaxies that will be observed by LSST.
We conclude in Section~\ref{sec:conclusions}.

\section{Data Set} \label{sec:data}

\subsection{Simulated LSST Data}

\begin{figure*}[t!]
    \centering
    \includegraphics[width=0.9\textwidth]{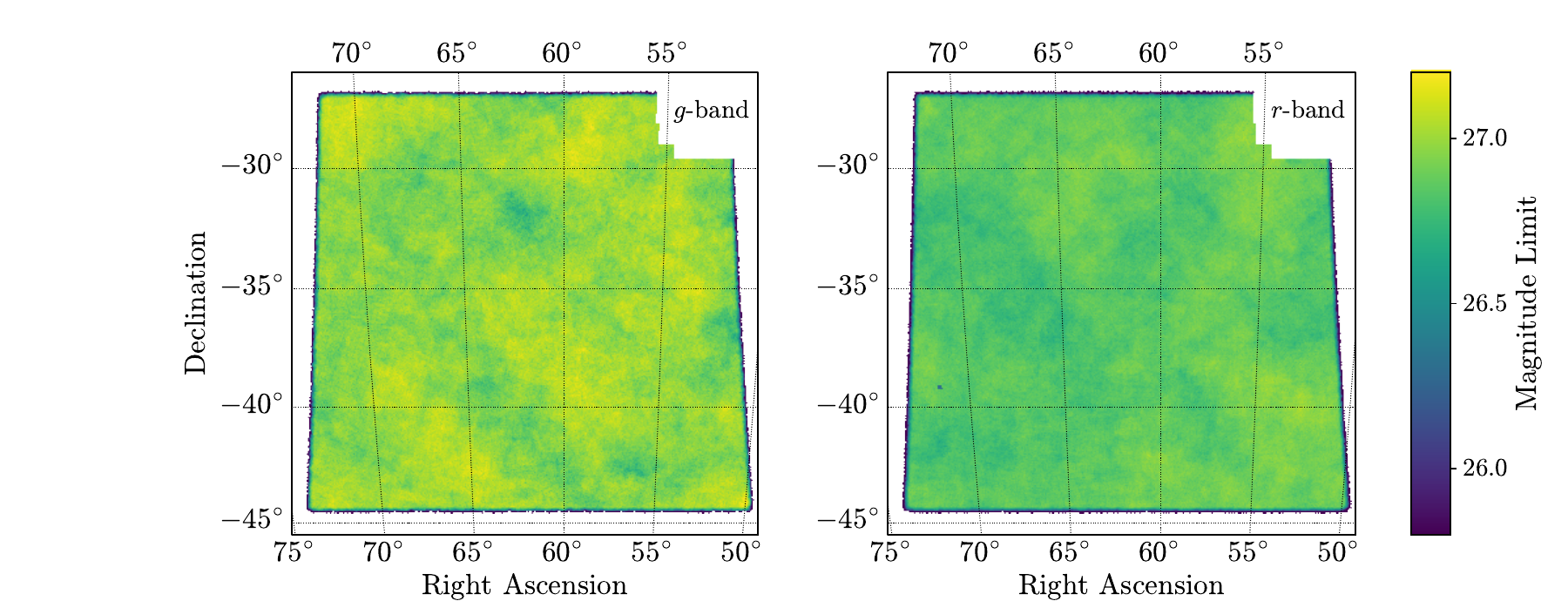}

    \caption{\label{fig:gr_depth_maps} The simulated LSST DESC DC2 footprint covers $\sim$300\,deg$^2$ of the high Galactic latitude sky. Maps of the ${\rm S/N} = 5$ magnitude limit for point-like sources in the $g$-band (left) and $r$-band (right) show the uniformity and depth of the WFD portion of DC2. Variations in the depth come from the the observing strategy and simulated observing conditions. The median ${\rm S/N} = 5$ magnitude limits for point-like sources are $g = 27.0$\,mag and $r = 26.8$\,mag.}
\end{figure*}

The starting point for our analysis is simulated data from LSST DESC DC2, an end-to-end simulated sky survey developed in preparation for extragalactic analyses of LSST data \citep[][]{Abolfathi_2021,lsstdarkenergysciencecollaboration2022desc}. The DC2 simulations aim to capture the characteristics of  high-Galactic-latitude LSST observations in a small (${\sim}\,300$\,deg$^2$) region centered on ${\rm RA}, {\rm Dec.} = 61.863^{\circ}, -35.790^{\circ}$ ($\ell, b = 237.34^\circ, -47.75^\circ$).
Input object samples were assembled from an N-body simulation of large-scale structure populated with galaxies \citep[i.e., the CosmoDC2 synthetic galaxy catalog based on the Outer Rim simulations;][]{Korytov:2019,Heitmann:2019} and a stellar population simulated to mimic the Milky Way bulge, disk, and halo \citep[i.e., based on \software{galfast};][]{Juric:2008}. 
Resolved stellar populations from Milky Way halo substructures (i.e., the Magellanic Clouds, other satellite galaxies, and stellar streams) were not included in the DC2 stellar catalog. The fluxes of simulated sources were reddened due to interstellar dust using the extinction maps from \citet{Schlegel:1998}; however, the effects of reddening are small due to the minimal extinction in the DC2 field, $E(B-V) \lesssim 0.07$.
DC2 consists of five years of simulated LSST observations generated with then-current estimates of the survey cadence and environmental conditions \citep[\texttt{minion\_1016};][]{Jones:2015}.  Realistic images were simulated by the \software{imSim} software,\footnote{\url{https://github.com/LSSTDESC/imSim/releases/tag/v0.6.2}} and processed with the LSST Science Pipelines to correct for simulated instrument signatures and produce object catalogs. The simulated DC2 sky survey includes six optical bands, \textit{ugrizy}, covering $\sim$\,300\,deg$^2$ of the LSST wide-fast-deep (WFD) footprint, as well as a deep drilling field (DDF) of approximately 1\,deg$^2$. The median ${\rm S/N} = 5$ magnitude limits for point-like sources in the DC2 WFD simulations are $g = 27.0$\,mag and $r = 26.8$\,mag (Figure~\ref{fig:gr_depth_maps}). The magnitude limit is estimated from the properties of the simulated DC2 observations (e.g., focal plane geometry, exposure time, image quality, and sky brightness) using the \texttt{supr\^{e}me} toolkit.\footnote{\url{https://github.com/LSSTDESC/supreme}}

The DC2 simulations have no throughput variation over the focal plane or between visits, so joint photometric and astrometric calibration across visits is not performed. Consequently, the standard passbands for the LSST filters are simply the total throughputs used in the simulations and were derived by the Rubin systems engineering team.\footnote{\url{https://github.com/lsst/throughputs/tree/DC2production}} 
We used these same throughputs to generate the stellar populations of simulated dwarf galaxies as described in Section \ref{subsec:sims}. 
Furthermore, DC2 does not include bright stars ($r \lesssim 10$\,mag), regions of high stellar density, or large interstellar extinction.
To mitigate the impact of these limitations, we mask regions of the LSST footprint where the survey performance is expected to differ appreciably from DC2 when making predictions for the observable population of Milky Way satellite galaxies (Section~\ref{sec:lum_func}).

\subsection{Data Access}

We accessed and analyzed the DC2 data through the RSP hosted on the Interim Data Facility (IDF) in the context of the first data preview (DP0.1) released in the period leading up to the start of LSST \citep{Guy:2023}. The goals of DP0.1 were to serve as an early integration test of the LSST Science Pipelines with the RSP, and to enable members of the scientific community to begin early preparations for science with LSST. DP0.1 was hosted on the IDF and access was provided through the RSP, which includes a set of integrated web-based applications, services, and tools to query, visualize, subset, and analyze LSST data \citep{RSP_ADASS2021}. The $\sim$300\,deg$^2$ WFD data set from DC2 was ingested and adopted as the primary data set for DP0.1.\footnote{\url{https://dp0-1.lsst.io/data-products-dp0-1}} In addition to the simulated output catalogs, DP0.1 contains a truth matched catalog to access properties of the astronomical objects used as input to DC2.  All analyses presented in this paper were performed using the RSP on the IDF, and a discussion of how we optimized our analysis to run on the RSP is provided in Appendix~\ref{app:rsp}. 

\subsection{Catalog Characterization}
\label{sec:character}

Following the procedure described in \citet{Drlica-Wagner:2020}, we characterized the key photometric properties of the DC2 catalog (i.e., photometric uncertainty, detection probability, and star/galaxy classification efficiency as a function of magnitude) to generate consistent catalog-level simulations of Milky Way satellites that can be combined with the DC2 object catalogs (Section~\ref{subsec:sims}).

\begin{figure*}[t!]
    \centering
    \includegraphics[width=\textwidth]{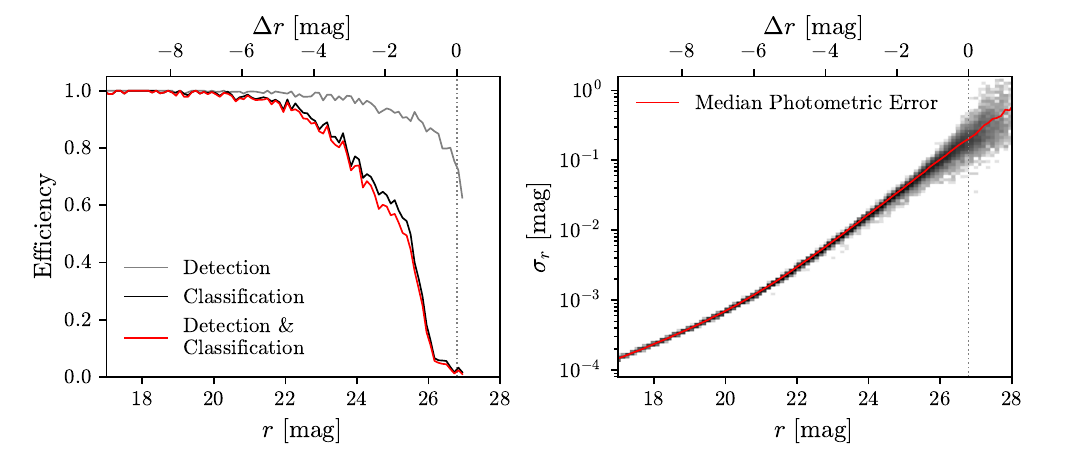}
    \caption{Characteristics of the simulated stellar catalog in a ${\sim}3$\,deg$^2$ region of DC2. We show the photometric performance as a function of the true $r$-band magnitude of simulated stars and the difference with respect to the $r$-band magnitude limit, $\Delta r = r - {\rm maglim}(r)$. (Left) Detection efficiency and star/galaxy separation efficiency models based on the input truth and measured output from DC2. 
    Star/galaxy classification is performed using the \texttt{EXTENDEDNESS} parameter output by the LSST Science Pipelines. Note the steep drop in stellar classification efficiency at $r \sim 26$ mag. 
    (Right) Photometric uncertainty of stars as a function limiting magnitude. The red line shows the median photometric uncertainty as a function of $r$, while the gray histogram shows the full distribution. This model is used to assign photometric uncertainties to simulated satellite member stars. In both panels, the gray dotted line indicates the $r$-band magnitude limit of DC2 at ${\rm S/N} = 5$ ($r = 26.8$\,mag).}
    \label{fig:efficiency-photerr}
\end{figure*}

To quantify star/galaxy separation efficiency, we selected a $\sim$3\,deg$^2$ region of the DC2 WFD footprint centered on $({\rm R.A., Dec.}) = (62\,\deg, -37\,\deg)$. Within this region, we selected a set of well-measured objects with truth matches and $g$- and $r$-band magnitude error, $\sigma_{g,r} < 0.2$ (S/N $\gtrsim 5$). We performed a simple star/galaxy separation based on the measured \texttt{EXTENDEDNESS} parameter \citep{Bosch:2018, lsstdarkenergysciencecollaboration2022desc}. The \texttt{EXTENDEDNESS} is a boolean classification variable defined from the difference between the point spread function (PSF) magnitude, and the CModel magnitude for objects. Point-like sources are have $\texttt{EXTENDEDNESS} = 0$, while extended galaxies have $\texttt{EXTENDEDNESS} = 1$ \citep{lsstdarkenergysciencecollaboration2022desc}.
To characterize the detection efficiency, we also tracked all stars in the region that were truth matched regardless of whether they passed our $\sigma_{g,r} < 0.2$ threshold. We binned these populations in true $r$-band magnitude to derive the stellar detection efficiency, stellar classification efficiency, and the simultaneous combination of the two (Figure~\ref{fig:efficiency-photerr}). We find a steep drop in stellar classification efficiency by $r \sim 26$\,mag.

To derive a model for the photometric uncertainty as a function of object magnitude and survey depth, we used the measured magnitude uncertainties of stars in combination with ${\rm S/N} = 5$ magnitude limit depth maps generated from the input images (Figure~\ref{fig:gr_depth_maps}). The magnitude limit maps were generated from the simulated observing properties (i.e., exposure time, exposure geometry, PSF FWHM, sky brightness) of the DC2 simulations using the \software{supr\^{e}me} package.\footnote{\url{https://github.com/LSSTDESC/supreme/}} We compute the median of the logarithm of the $r$-band magnitude uncertainty as a function of the difference between the true input $r$-band magnitude of the simulated object and the value of the magnitude limit map at the location of the object, $\Delta r = r - {\rm maglim}(r)$ (Figure~\ref{fig:efficiency-photerr}). This model for the expected magnitude uncertainty as a function of magnitude and survey magnitude limit is applied to the simulated satellite member stars to generate realistic photometric scatter in magnitude and color (Section~\ref{subsec:sims}).

\begin{figure*}[t!]
    \hspace{1.4cm} \includegraphics[width=\textwidth]{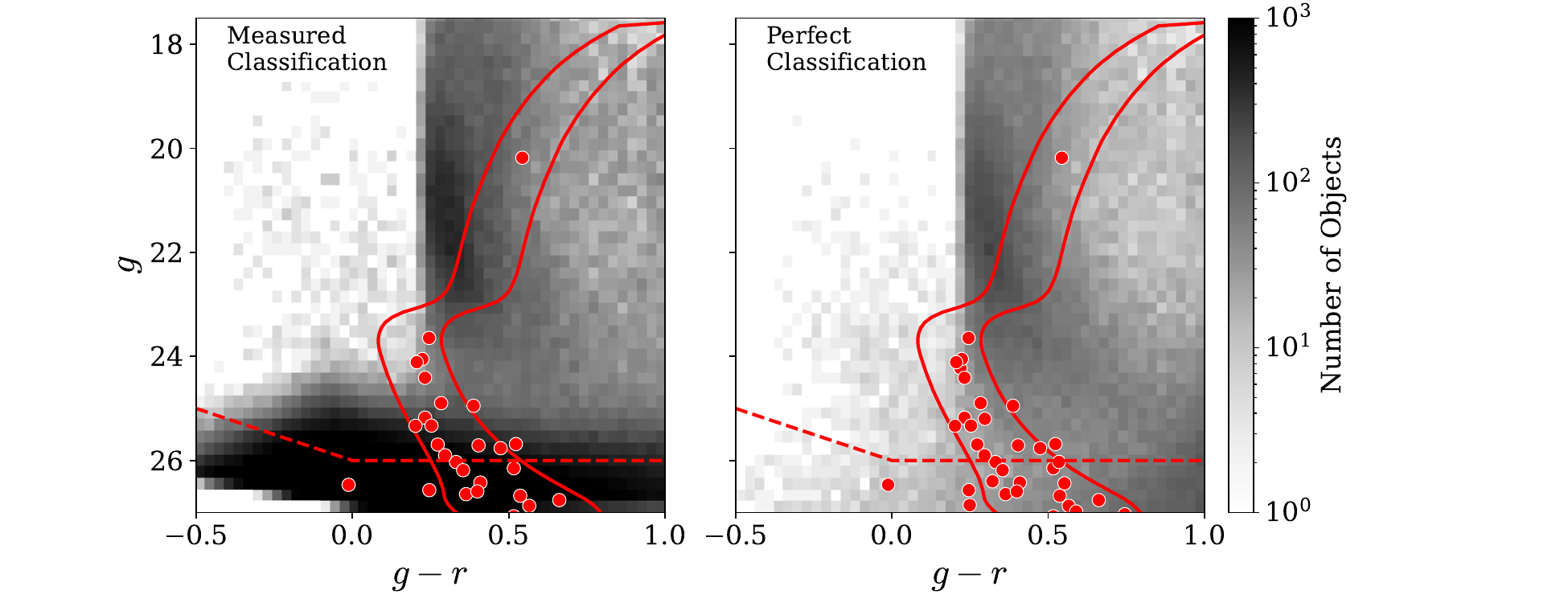}
    \caption{\label{fig:cmd_ideal_measured} Simulated LSST color--magnitude diagram of stars associated with a simulated Milky Way satellite (red points; $M_V = -0.3$\,mag, $D = 91$\,kpc) and the density of simulated objects from DESC DC2 that are classified as stars (gray background). The left panel uses a measured star/galaxy classification based on the $\texttt{EXTENDEDNESS}$ parameter while the right panel assumes a perfect star/galaxy separation. The red solid line shows our isochrone selection region, and the dashed red line indicates the magnitude cut at $g < 26$\,mag and $r < 26$\,mag that is applied in our analysis. The population of faint ($g \gtrsim  25$\,mag), blue ($g-r \lesssim 0.25$) background galaxies that are mis-classified as stars presents a significant contamination in searches for faint Milky Way satellites. We note that the sharp break in stellar density at $g - r \sim 0.25$ comes from the main sequence turn-off of simulated Milky Way halo stars in DC2.}
\end{figure*}

\subsection{Simulated Satellites} 
\label{subsec:sims}

We generated artificial Milky Way satellites as collections of individually resolved stars to determine the detection efficiency of our isochrone matched-filter search. We randomly sampled values for each satellite's stellar luminosity, heliocentric distance, physical size, ellipticity, position angle, age, and metallicity (Table~\ref{tab:sim_params)}). These parameters were selected to broadly represent the characteristics of the known population of Milky Way satellites \citep[e.g.,][]{Pace:2024}.
 Generating satellites over a wide range of parameter space ensures that the detection efficiency can be calculated in an unbiased manner. 
The satellites were spread across the DC2 footprint, with some satellites placed slightly outside the boundary of the footprint.

Satellites were simulated at the catalog level as a collection of individually resolved stars. 
This assumption is motivated by the range of physical sizes and distances of the satellites that we simulate, along with expectations for the FWHM of the LSST PSF.
It has been found that full image simulations are necessary to capture the effects of blending on compact systems at larger distances \citep[e.g.,][]{Mutlu-Pakdil:2021, Zhang:2025}.
We simulated stellar catalogs using estimates for the object detection completeness, star/galaxy separation, and photometric uncertainty derived from the DC2 catalogs (Section~\ref{sec:character}; Figure \ref{fig:efficiency-photerr}). To maintain realism, a probabilistic model was used for the spatial and flux distributions of stars within a satellite. Spatial distributions were sampled from a Plummer profile \citep{10.1093/mnras/71.5.460}, which is a good description of the stellar distribution of observed Milky Way satellite galaxies \citep{Simon:2019}. In this work, we use $a_h$ to indicate the angular elliptical semi-major axis containing half the light (arcmin) and $r_h = a_h\sqrt{1 - e}$ to indicate the azimuthally averaged angular half-light radius (arcmin), where $e$ is ellipticity. The corresponding physical 2D sizes (pc) at heliocentric distance, $D$, are indicated with $a_{1/2}$ and $r_{1/2}$, respectively.

The simulated LSST bandpasses used for DC2 were integrated into the Padova CMD interface to generate photometry consistent with DC2.\footnote{\url{http://stev.oapd.inaf.it/cgi-bin/cmd}} The initial masses of satellite member stars were drawn from a \citet{Chabrier_2001} initial mass function (IMF),  which has been found to be a reasonable description of known satellite galaxies  
\citep{Simon:2019}. These initial masses were used to assign current absolute magnitudes from a \citet{Marigo:2017} isochrone. When sampling from the IMF, the lower mass bound was set to the hydrogen-burning limit of $0.08 M_{\odot}$ and the upper bound was set by the star with the largest initial mass in the evolved isochrone (post-asymptotic giant branch and white dwarf stars are excluded). Using the \citet{Marigo:2017} isochrones, we transformed from initial stellar mass to current absolute magnitude in the $g$ and $r$ bands of the DC2 photometric system, and then to apparent magnitudes using the distance modulus of the simulated satellite. Detectability, stellar classification probability, photometric scatter, and photometric measurement uncertainties were assigned to each simulated star based on its true apparent magnitude and the DC2 survey limiting magnitude at the location that the star was injected following the procedures described in Section~\ref{sec:character}. In this way, our simulated stellar catalog mimics the DC2 catalog in detection efficiency, object (mis-)classification, and photometric uncertainty, and thus it can be combined with the DC2 catalog directly.

\begin{center}
\begin{deluxetable}{c c c c }
\tablecaption{Parameter ranges for simulated satellites
\label{tab:sim_params)}}
\tablehead{
\colhead{Parameter} & \colhead{Range} & \colhead{Sampling} & \colhead{Unit}
}
\startdata
Distance                   & [5, 500]                             & log     & kpc\\
Stellar mass               & [10, $10^6$]                         & log     & $M_\odot$ \\
2D half-light radius       & [1, 2000]                       & log     & pc\\
Ellipticity                & [0.1, 0.7]                           & linear  & \ldots \\
Position Angle             & [0, 180]                             & linear  & deg \\
Metallicity                & \{$1\times10^{-4}$, $2\times10^{-4}$\} & choice & $Z/Z_\odot$ \\
Age                        & \{10, 12, 13.5\}                      & choice & Gyr \\[-0.5em]
\enddata 
\tablecomments{Distance, stellar mass and physical half-light radius are drawn from uniform distributions in log space. Ellipticity and position angle are drawn from a uniform distribution, while age and metallicity are discrete choices. Isochrones were generated using the models of \citet{Marigo:2017} with an IMF from \citet{Chabrier_2001}.}
\end{deluxetable}
\end{center}

\section{Analysis} 
\label{sec:analysis}

This section provides a general overview of the analysis pipeline used to derive an observational selection function for Milky Way satellite searches using LSST. 
The search was performed on $10^5$ simulated satellites. 
For each simulated satellite, we queried the Rubin Table Access Protocol (TAP) service for catalog objects that were located $\leq 2$\,deg from the centroid of the simulated satellite. We injected the simulated satellite into the DC2 region and restricted the merged catalog to objects with magnitude error in the $g$ and $r$ bands of $\sigma_{g,r} < 0.2$ (${\rm S/N} \gtrsim 5$). Furthermore, we imposed a magnitude cut of $g < 26$\,mag and $r < 26$\,mag on both our simulated satellites and the DC2 data. 
This cut was imposed due to the sharp drop in stellar classification efficiency at $r \sim 26$\,mag (Figure~\ref{fig:efficiency-photerr}). Extending to fainter magnitudes resulted in higher contamination from mis-classified galaxies and a higher false positive rate for our isochrone matched-filter search (Section~\ref{sec:results}).

To explore the impact of star/galaxy classification efficiency, we ran the analysis in two different configurations. In the first, the star/galaxy classification was performed using the measured $\texttt{EXTENDEDNESS}$ parameter: objects with $\texttt{EXTENDEDNESS} = 0$ were classified as stars, and those with $\texttt{EXTENDEDNESS} = 1$ were classified as background galaxies. In the second configuration, the star/galaxy classification instead used the true object class from the simulation inputs. The difference in star/galaxy classification efficiency was also applied when generating simulated satellites---i.e., we generated two separate sets of satellite simulations applying the measured and perfect star/galaxy classification efficiency.
Example color--magnitude diagrams of the objects passing these selections for one simulated satellite are shown in Figure~\ref{fig:cmd_ideal_measured}.
Downstream analysis procedures were the same for these two different star/galaxy separation procedures. 
We note that the sharp break in the stellar density at $g - r \sim 0.25$ (Figure~\ref{fig:cmd_ideal_measured}) comes from the main sequence turn-off of Milky Way halo stars in DC2 generated by \texttt{galfast} \citep{Juric:2008}. Similar sharp features are seen in deep photometric data \citep[e.g.,][]{Pieres:2020} and other simulations of the LSST stellar population \citep[e.g.,][]{DalTio:2022}.

No interstellar extinction corrections were applied to the DC2 data in this analysis. The DC2 footprint is located at high Galactic latitude where reddening is minimal, and we find that applying extinction corrections to the data resulted in a negligible (${\sim}$0.25\%) change in the significance at which simulated satellites are detected.

\subsection{Search Algorithm} \label{sec:simple}

Our automated, matched-filter search was implemented as the \simple algorithm developed for satellite searches in the Dark Energy Survey \citep[DES;][]{Bechtol:2015, Drlica-Wagner:2020}.\footnote{\url{https://github.com/sidneymau/simple_adl/tree/kb}} This algorithm works by applying an isochrone filter in color--magnitude space to enhance the contrast of an old, metal-poor stellar system relative to a local estimate of Milky Way foreground stars and mis-classified background galaxies (Figure~\ref{fig:cmd_ideal_measured}). The filtered stellar density field was smoothed by a Gaussian kernel ($\sigma = 2$\,arcmin), and we identified local density peaks by iteratively raising a density threshold until there are fewer than 10 disconnected regions above the threshold value. In practice, only the most prominent of these stellar overdensities passed our minimal statistical significance thresholds. At the central location of each density peak, \simple determines the angular size of a surrounding circular aperture that maximizes the significance of the density peak with respect to the distribution of field stars. This is done by iterating through apertures with radii ranging from 0.01\,deg to 0.30\,deg. \simple computes the characteristic local density by counting stars in an annulus from 0.3\,deg to 0.5\,deg. This characteristic local density is then used in combination with the aperture size to predict the expected number of stars in the source field. The actual observed counts are then calculated. The primary output is a list of candidates and associated Poisson detection significance, SIG, which ranges between $0 \leq {\rm SIG} \leq 37.5$. The upper limit here corresponds to a numerical limit of the inverse survival function of the normal distribution in \software{scipy}, which is itself calculated from the multivariate survival function of the expected number of stars and number of observed stars. This process was repeated for every computed aperture, and the maximum SIG value is reported as the detection significance.

To estimate the LSST observational selection function, we ran the search on regions of interest centered on each injected satellite.
Furthermore, we fixed the distance modulus of the search to the true distance modulus of the injected satellite rather than scanning over distance moduli.  This approach significantly reduced the computational cost to a level that is easily achievable on the IDF (see Appendix \ref{app:rsp}).
Past studies have found that these simplifying choices introduce trivial changes in the measured significance relative to what would be estimated from a rigorous uninformed scan in distance modulus \citep{Drlica-Wagner:2020}.
The broad isochrone selection performed by \simple was centered around a \citet{2012MNRAS.427..127B} isochrone with an age of 12 Gyr and metallicity of $Z = 0.0002$. The apparent $g$- and $r$-band magnitudes for this isochrone were calculated in the DES filter bandpasses \citep{DES:2018}. We verify that the inconsistency between the isochrone and bandpasses used for the satellite simulations (Section~\ref{subsec:sims}) and the search has a negligible effect (${\sim}0.5\%$ change in detection significance) due to the broad selection that \simple applies around this isochrone (Figure~\ref{fig:cmd_ideal_measured}).
For each of the $10^5$ injected satellites, we store the detection significance along with the right ascension, declination, distance modulus of the most significant peak. We also store the angular size of the circular aperture that maximizes the detection significance, the number of stars observed within that aperture, and the number of stars expected within the aperture given the characteristic local density estimate. Finally, we store the number of stars observed within a best-estimate of the half-light radius.

Our light-weight implementation of the \simple algorithm does not incorporate knowledge of the observational coverage of the DC2 footprint. This leads to false-positive detections on the edges of the DC2 footprint where the catalog density changes abruptly. These edge cases were filtered out of this analysis, since they will be dealt with more rigorously in the analysis of real data \citep[e.g.,][]{Drlica-Wagner:2020}.
Satellites that are very large (having an angular size $r_h \gtrsim 2$\,deg) have a low detection significance because the algorithm struggles to differentiate between the source and background stellar populations, since the entire aperture is filled with satellite member stars.
These large systems have low detection probability in our analysis, but large and moderately bright systems have been discovered through other search techniques.\footnote{Proper motions and variable stars are powerful observational tools to discover extended, low-surface-brightness systems, as demonstrated in the discovery of the Antlia~II satellite \citep[$r_h \gtrsim 1$\,deg;][]{Torrealba:2019}.}

\begin{figure}[t!]
    \centering
    \includegraphics[width=\columnwidth, trim=0cm 0cm 1cm 0cm, clip]{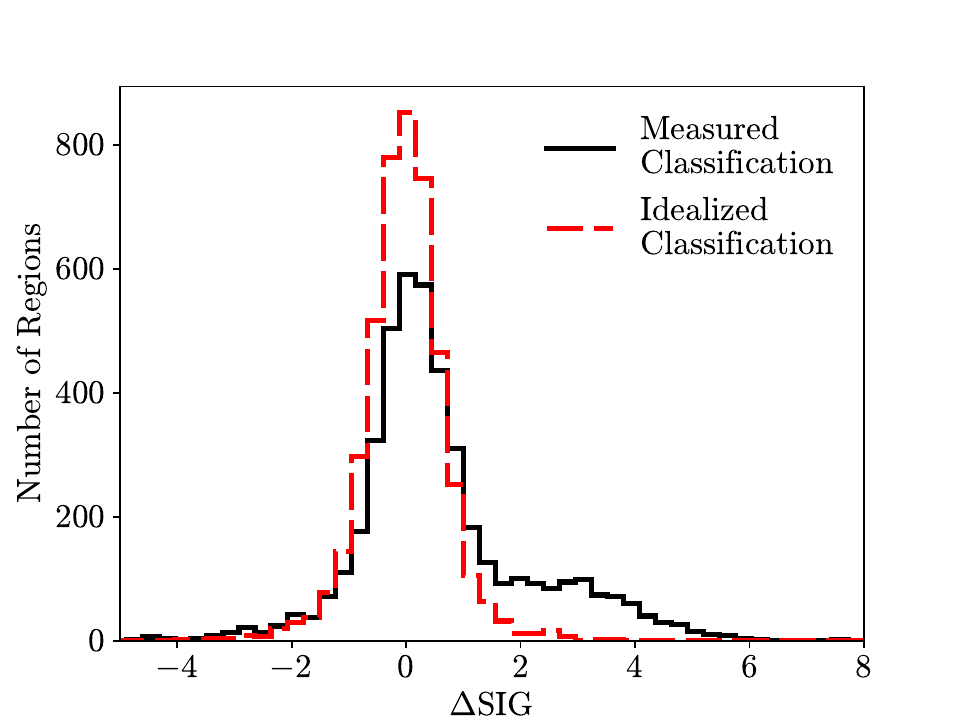}
    \caption{\label{fig:bmodel} Spatial clustering in the distribution of mis-classified galaxies can increase the rate of false positives in Milky Way satellite searches. We show how the detection significance of blank-sky regions changes when the spatial positions of DC2 objects classified as stars are randomly shuffled, $\Delta {\rm SIG} = {\rm SIG_{\rm DC2} - SIG_{\rm rand}}$. We find a noticeable tail to larger $\Delta {\rm SIG}$ values when we use the measured star/galaxy separation (black solid line) compared to idealized star/galaxy separation (red dashed line).} 
\end{figure}

\subsection{Foreground/Background Model} \label{sec:bmodel}

The \simple algorithm uses an outer annulus to model the distribution of foreground and background objects. 
It interpolates this background model to the region of interest assuming a uniform spatial distribution and no variation in the color--magnitude distribution of objects (Section \ref{sec:simple}).
Interpreting the detection significance as a Gaussian $p$-value to yield the chance probability of a false positive detection relies on the assumption that our uniform background model is the correct model for the data.
Deviations from this assumption (i.e., spatial or photometric structure in the background) can bias our results and lead to an increased false positive rate.  

To test the validity of the background model, we performed a null test where we ran the search on 1000 randomly selected regions of the DC2 field without injecting any simulated satellite. For each region, we ran \simple and stored the most significant detection. We then  randomized the positions of the DC2 objects and reran \simple, storing the most significant detection in the randomized field. We calculated the change in significance, $\Delta {\rm SIG} = {\rm SIG_{\rm orig} - SIG_{\rm rand}}$, where we expect $\Delta {\rm SIG} = 0$ on average if the DC2 background is uniform. This process was done separately for the analyses using measured star/galaxy separation and perfect star/galaxy separation.

We show the results of our background null tests in Figure~\ref{fig:bmodel}. We find that with perfect star/galaxy separation, there is little bias toward positive or negative changes in the significance when the DC2 objects are randomized.  However, when using measured star/galaxy separation, we find a noticeable bias towards higher significance with the non-randomized background. This suggests that the measured star/galaxy separation is introducing a tendency to measure higher significances in blank-sky regions than would be expected from a uniform background field with the same photometric information. One possible explanation for this result is that the leakage of mis-classified galaxies into the stellar sample may introduce a spatially structured background due to the large-scale structure of galaxies. These results are discussed further in Section \ref{sec:results}.

\begin{figure*}[t!]
    \centering
    \includegraphics[width=1.0\textwidth]{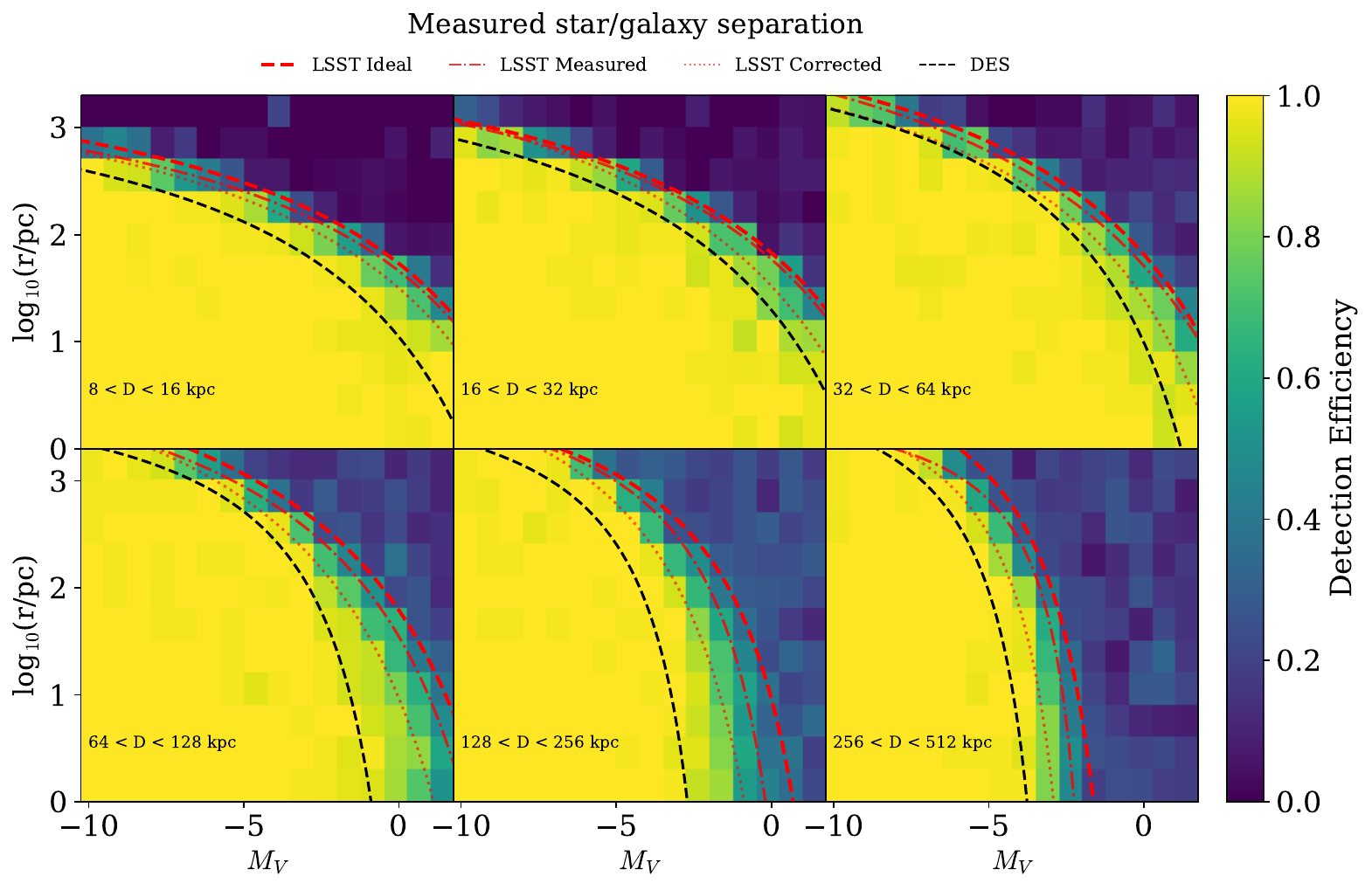}
    \caption{Detection efficiency of searches for simulated Milky Way satellites in DC2 using measured star/galaxy information. Detection efficiency is reported as a function of azimuthally averaged physical half-light radius, and absolute V-band magnitude in six different bins of heliocentric distance (logarithmically spaced from 8 kpc to 512 kpc). Detection efficiency ranges from $0\%$ (purple) to $100\%$ (yellow). Overplotted are 50\% detection efficiency contours for: DES (black dashed line), measured star/galaxy classification with an adjusted detection threshold of ${\rm SIG} > 8.4$ (red dotted line), measured star/galaxy classification with a threshold of ${\rm SIG} > 5.5$ (red dot-dash line) and perfect star/galaxy classification with a threshold of ${\rm SIG} > 5.5$  (red dashed line). The greatest gains relative to previous surveys are found for faint and relatively compact objects at larger distances ($D \geq 64$\,kpc).}
    \label{fig:measured_osf}
\end{figure*}

\begin{figure*}[t!]
    \centering
    \includegraphics[width=1.0\textwidth]{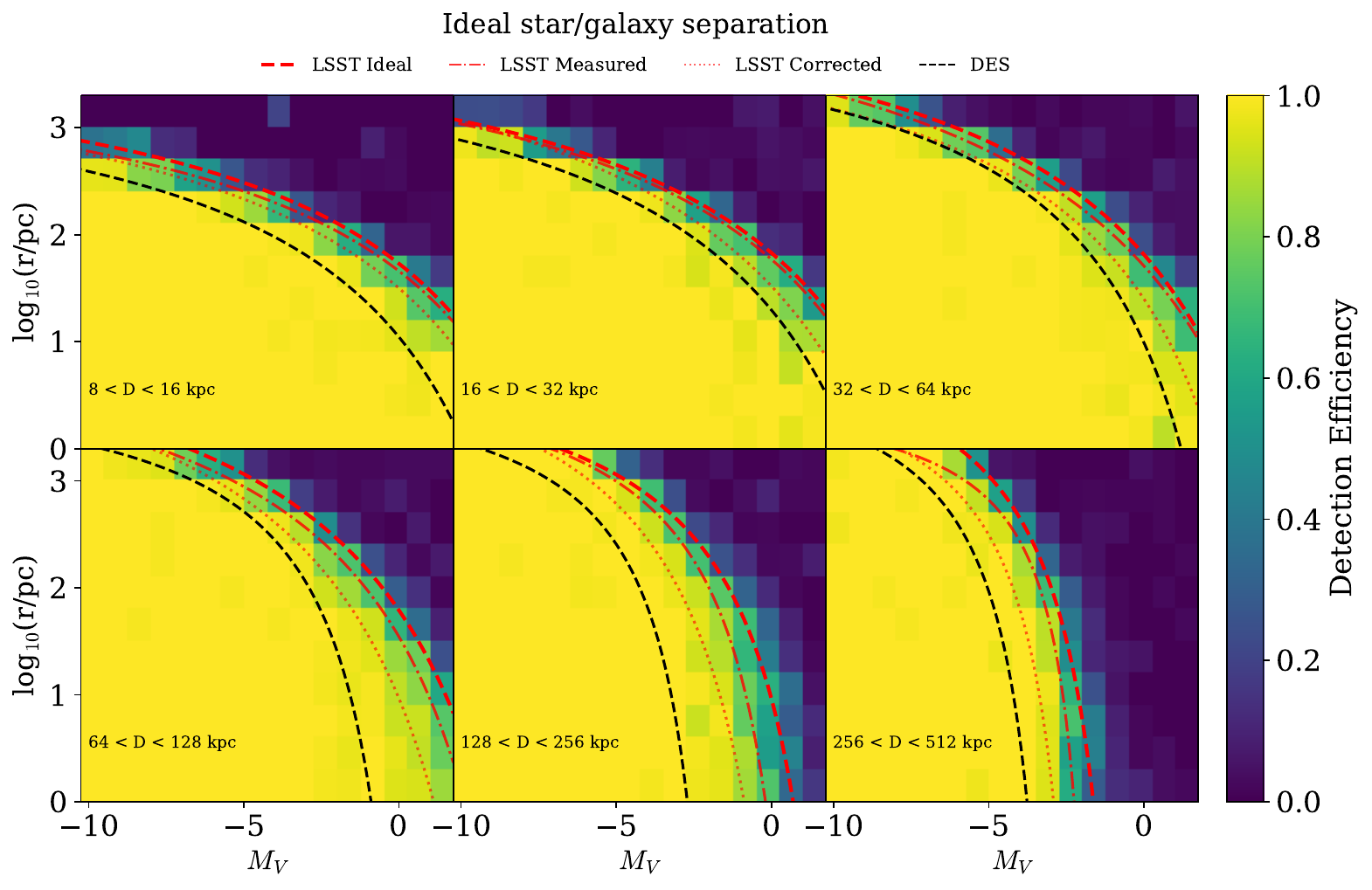}
    \caption{Similar to Figure~\ref{fig:measured_osf}, but showing observational selection function assuming perfect star/galaxy classification that uses the true object class. Overplotted are 50\% detection efficiency contours for: DES (black dashed line), measured star/galaxy classification with an adjusted detection threshold of ${\rm SIG} > 8.4$ (red dotted line), measured star/galaxy classification with a threshold of ${\rm SIG} > 5.5$ (red dot-dash line) and perfect star/galaxy classification with a threshold of ${\rm SIG} > 5.5$  (red dashed line). The largest impact of the star/galaxy classification efficiency occurs for faint satellites at larger distances.}
    \label{fig:ideal_osf}
\end{figure*}

\section{Observational Selection Function} 
\label{sec:osf}

We present the results of our DC2 analysis in the form of an observational selection function that describes the detectability of a satellite as a function of its physical properties \citep[e.g.,][]{Koposov_2008, Walsh_2009, Drlica-Wagner:2020}{}{}. The observation selection function describes the probability that a satellite will have a measured detection significance that is greater than a given threshold (i.e., ${\rm SIG} > 5.5$) as a function of its heliocentric distance ($D$), absolute $V$-band magnitude ($M_V$), and azimuthally averaged projected physical half-light radius ($r_{1/2}$). 
We present two different parameterizations of the observational selection function: an analytic approximation of the 50\% detection efficiency contour and a machine learning model describing the full range of detectability. 

\subsection{Effects of Star/Galaxy Classification}
\label{sec:results}

We define our canonical sample of detected objects as those that were recovered with ${\rm SIG} > 5.5$, consistent with the threshold imposed by \citet{Drlica-Wagner:2020} when analyzing the DES Y3 and Pan-STARRS1 data sets. \citet{Drlica-Wagner:2020} motivated the choice of this threshold value in two ways. Firstly, the observed number of unconfirmed candidates was found to increase rapidly if the threshold was reduced. Secondly, systems detected above this threshold could be unambiguously classified as genuine stellar systems or obvious survey artifacts \citep[see Appendix D of][]{Drlica-Wagner:2020}.  We plot our resulting selection functions in Figures~\ref{fig:measured_osf} and \ref{fig:ideal_osf}.  We overplot the parameterized 50\% detection efficiency contours (Section~\ref{sec:anapprox}) from our analyses and the DES Y3 search to illustrate the projected increase in satellite sensitivity due to increases in survey depth from LSST. 
While LSST is more sensitive at all distances, the gains are most appreciable at distances $D > 64\,{\rm kpc}$. 
We find that the difference between measured and perfect star/galaxy separation has a relatively minor impact on the 50\% detection efficiency contour, which is most significant for faint ($M_V \sim 0$\,mag), compact ($\log_{10}(r_{1/2}/{\rm pc}) \lesssim 1.0$), and distant ($D \gtrsim 128$\,kpc) systems.

While the 50\% detection efficiency contours are reasonably similar for the measured and perfect star/galaxy separation analyses, we note that the false positive rate (i.e., the fraction of blank fields that are detected with ${\rm SIG} > 5.5$) is appreciably higher with the measured star/galaxy separation. 
This can be seen by comparing the top right corners of Figures \ref{fig:measured_osf} and \ref{fig:ideal_osf}.
In this regime, the injected satellites are far too faint and diffuse to be detected by \simple.
However, ${\sim}\,20\%$ of these injected satellites are ``detected'' above the significance threshold when the measured star/galaxy classification is used (Figure~\ref{fig:measured_osf}).
Quantitatively, we find that the false positive rate is 2.3\% using perfect star/galaxy separation compared to 22.4\% using measured star/galaxy separation.
This is a direct result of the bias shown in Figure \ref{fig:bmodel}, which shows that using the measured star/galaxy separation results in a non-negligible bias toward high SIG values.
Interestingly, Figure~\ref{fig:measured_osf} shows that the false positive rate appears to depend on the heliocentric distance of the simulated satellite.
As discussed in Section \ref{sec:analysis}, we fixed the distance modulus of the \simple search at the true distance modulus of the injected satellite.
For satellites at intermediate and large distances, this leads to a larger overlap of the isochrone filter with the locus of background galaxy contamination in the color--magnitude diagram (Figure~\ref{fig:cmd_ideal_measured}).
Such contamination could be further mitigated by more advanced satellite search algorithms---e.g., algorithms that incorporate information about the expected IMF of stellar systems \citep[e.g.,][]{Bechtol:2015, Drlica-Wagner:2020}.

To perform a more rigorous comparison between the selection efficiencies assuming the measured and perfect star/galaxy classification, we adjusted the detection threshold for the analysis using the measured star/galaxy classification to match the false positive rate of the perfect star/galaxy classification analysis. This was achieved by raising the detection threshold from ${\rm SIG} > 5.5$ to ${\rm SIG} > 8.4$ when using the measured star/galaxy classification, yielding a false positive rate of 2.3\%. The resulting 50\% detection efficiency using the higher threshold (red dotted line in Figures \ref{fig:measured_osf} and \ref{fig:ideal_osf}) is noticeably shallower than the 50\% detection efficiency with the lower threshold (red dashed line), although both are appreciably deeper than DES (black dashed line). 
The drop in efficiency resulting from requiring a higher detection threshold emphasizes the importance of star/galaxy classification when it comes to searches for resolved stellar systems with LSST.

\subsection{Analytic Approximation}\label{sec:anapprox}

A simple analytic approximation of the 50\% detection efficiency contour is a sufficient description of the sensitivity of previous searches for some applications \citep{Koposov_2008, Walsh_2009, Drlica-Wagner:2020}. Following \citet{Drlica-Wagner:2020}, we parameterize the 50\% detection efficiency contour, $P_{\rm det,50}$, at fixed distance by:
\begin{equation}
\log_{10}(r_{1/2}) = \frac{A_0(D)}{M_V - M_{V,0}(D)} + \log_{10}(r_{1/2,0}(D))
\end{equation}
\noindent where $r_{1/2}$ is in units of pc, $M_V$ is in units of mag, and D is in units of kpc. We fit the three distance-dependent constants, $A_0$, $M_{V,0}$, and $r_{1/2,0}$,  in each of six heliocentric distance bins from 8\,kpc to 512\,kpc.  The constants $M_{V,0}$ and $r_{1/2,0}$ represent asymptotic limits in absolute magnitude and physical half-light radius as a function of satellite distance. The scale parameter, $A_0$, describes the ``radius of curvature'' of the $P_{\rm det,50}$ contour at a given distance. We plot the contours as the dashed lines in Figures \ref{fig:measured_osf} and \ref{fig:ideal_osf}, and we provide the values of the best-fit coefficients in Table 2. 

\subsection{Machine Learning Model}\label{sec:mlmodel}

While the analytic satellite sensitivity function described in Section \ref{sec:anapprox} has been found to be a sufficient description of the search sensitivity for some applications, it does not fully characterize the intermediate region between  0\% and 100\% detection efficiency. This is overcome by characterizing the sensitivity of our search algorithm with a machine learning algorithm that learns the full behavior of the observational selection function\footnote{\url{https://github.com/LSSTDESC/dc2_satellite_census}}.

Following the procedure described in Section 7.2 of \citet{Drlica-Wagner:2020}, we trained a gradient-boosted decision tree classifier to predict the probability of detecting a satellite as a function of its physical properties: absolute $V$-band magnitude, azimuthally averaged projected physical half-light radius, and distance. This is a binary classification problem, where we predict the relationship between a set of input features, $\vec{X}$, and a set of labels, $\vec{Y}$. Each simulated satellite represents a training instance, $i$, with its physical properties comprising the feature vector, $\vec{X_i}$. Satellites are labelled as detected ($Y_i = 1$) or undetected ($Y_i = 0$) based on the detection criteria described in Section~\ref{sec:results}.  Our classifier was then trained to output the probability that a satellite would be detected.

Our gradient-boosted decision tree classifier was trained using \software{XGBoost} \citep{10.1145/2939672.2939785} and \software{scikit-learn} \citep{pedregosa2018scikitlearn} as follows:

\begin{enumerate}[nosep]
\item Randomly split the simulated satellites into training and test sets that contain 90\% and 10\% of the simulated satellites, respectively.

\item Randomly split the training set from the previous step into $k$ hold-out cross-validation subsets. We chose $k = 3$ for this analysis by performing a manual grid search over different numbers of cross-validation folds.

\item Train \texttt{XGBClassifier} using \texttt{GridSearchCV} to select hyperparameters that produce the best test-set classification score. Hyperparameters include the learning rate, the number of trees, and the maximum tree depth
\end{enumerate}

The trained \software{XGBoost} model can be used to predict the probability that a satellite will be detected as a function of its physical parameters. We proceed to use this model to predict the number of Milky Way satellite galaxies that LSST will observe based on a cosmological model of the Milky Way satellite galaxy population.

\begin{deluxetable}{c c c c }
\tablecaption{\label{tab:p50_params} Parameterization of the 50\% Detection Efficiency Contours for LSST.}
\tablehead{Distance & $A_0$ & $M_{V,0}$ & $\log_{10}(r_{1/2,0}/{\rm pc})$ \\
(kpc) & & (mag) &  }
\startdata
\multicolumn{4}{c}{Idealized Star/Galaxy (${\rm SIG}>5.5$)} \\
\hline
11.3 &  22.7 & 10.0 & 4.0 \\
22.6 &  24.7 & 10.0 & 4.3 \\
45.2 &  25.1 & 8.7  & 4.7  \\
90.5 &  24.4 & 7.6  & 5.0  \\
181  &  11.0 & 3.2  & 4.4 \\
362  &  8.6  & 0.2  & 4.7  \\
\hline
\multicolumn{4}{c}{Measured Star/Galaxy (${\rm SIG}>5.5$)} \\
\hline
11.3 &  22.4 &  10.0 & 3.9  \\
22.6 &  25.3 &  10.0 & 4.3  \\
45.2 &  31.6 &  9.9  & 4.9 \\
90.5 &  21.5 &  6.6  & 4.8 \\
181  &  9.0  &  1.9  & 4.3  \\
362  &  4.1  &  -1.2 & 3.9 \\
\hline
\multicolumn{4}{c}{Measured Star/Galaxy (${\rm SIG}>8.4$)}\\
\hline
11.3 & 25.0  & 10.0  &  4.0 \\
22.6 &  30.8 & 10.0  & 4.6  \\
45.2 &  20.7 &  6.9 &  4.4 \\
90.5 &  23.4 & 5.8  & 5.0  \\
181  &  14.3 &  2.1 &  4.8 \\
362  &  6.5 &  -1.4 & 4.3  \\[-1.0em]
\enddata
\end{deluxetable}

\begin{figure*}[t!]
    \hspace{1.75cm}\includegraphics[width=0.9\textwidth]{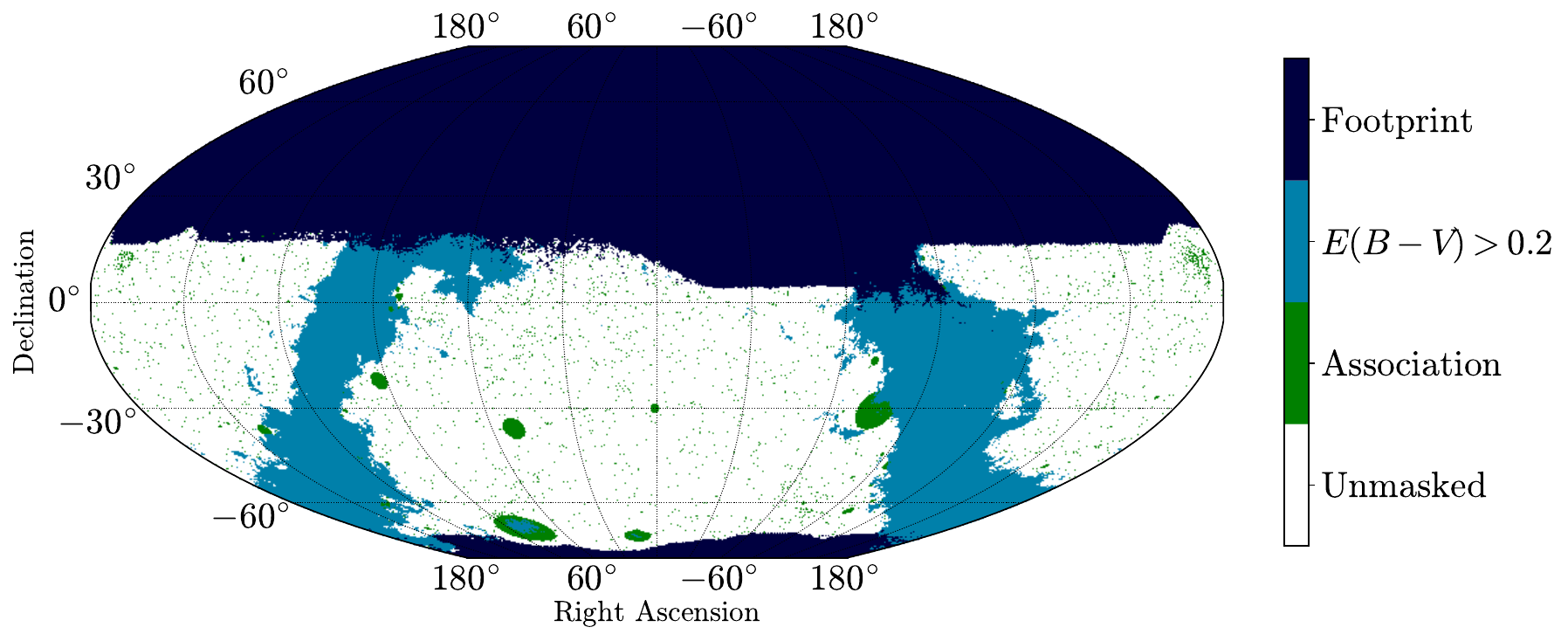}
    \caption{\label{fig:mask} Geometric masks applied to predict the population of Milk Way satellite galaxies that will be observable by LSST. Colored regions are removed from our analysis either because they lie outside the LSST WFD footprint (dark blue), have interstellar extinction $E(B-V) > 0.2$ in the \citet{Schlegel:1998} dust map (light blue), or are associated with previously known stellar overdensities, galaxy clusters, or bright stars (green). The total unmasked area is $\sim$18,300\,deg$^2$.}
\end{figure*}

\section{Satellite Galaxy Luminosity Function} 
\label{sec:lum_func}

To predict the population of Milky Way satellite galaxies that will be be observed by LSST, we combine the observational selection functions derived from our studies of DC2 (Section~\ref{sec:osf}), predictions for the depth and coverage of the 10-year LSST WFD program, and a cosmological model of the Milky Way satellite population \citep{Nadler:2020}.\footnote{\url{https://github.com/eonadler/subhalo_satellite_connection}} We focus on predicting the number of satellite galaxies with $M_V < 0$\,mag and $r_{1/2} > 10$\,pc that LSST will detect within the virial radius of the Milky Way halo (assumed to be 300\,kpc) as a function of satellite luminosity. 
Note that our predicted luminosity function assumes an anisotropic distribution of satellite galaxies. The  galaxy--halo model inference was conditioned on the presence of the Large Magellanic Cloud (LMC), with the sky position and distance of the LMC fixed to its measured value, such that it lies at the correct angular position relative to the LSST footprint \citep{Nadler:2020}.
In contrast to some previous analyses \citep{Drlica-Wagner:2020, Nadler:2020}, we neglected an explicit dependence of search sensitivity on stellar density, and thus on angular position within the LSST footprint. The DC2 footprint is located at high Galactic latitude, giving us little ability to assess the impact of variations in stellar density on the observational sensitivity of LSST. Furthermore, photometry in more crowded fields is an area of ongoing development in the LSST Science Pipelines.

We estimate the coverage and depth of the 10-year LSST WFD program using the \texttt{baseline} v4.0 operations simulation.\footnote{\url{https://survey-strategy.lsst.io/baseline/index.html}} This model reflects the current understanding of the Rubin system throughput and LSST observing strategy, and it represents a significant update from the operations simulation used to generate DC2 (\texttt{minion\_1016}; \citealt{Jones:2015}). We compared the measured depth of the DC2 simulations to the WFD region of the \texttt{baseline} v4.0 survey, and we find comparable median depths in the $g$- and $r$-bands. Thus, we restrict the \texttt{baseline} v4.0 footprint to regions where the ${\rm S/N} = 5$ point-source depth in the $g$- and $r$-band depth is $\geq 26$\,mag to reflect the cut that was applied on our DC2 satellite analysis (Section~\ref{sec:analysis}). In addition, we masked regions of the sky where we expect Milky Way satellite discovery to be complicated due to interstellar dust, $E(B-V) > 0.2$ \citep{Schlegel:1998}, or close associations with other astronomical systems. 
Following Section 6.1 of \citet{Drlica-Wagner:2020}, these astronomical associations include regions around bright stars \citep{1991bsc..book.....H}, globular clusters of the Milky Way and Magellanic Clouds (\citealt{Harris:1996}, 2010 edition; \citealt{2008MNRAS.389..678B}), open clusters (WEBDA)\footnote{\url{https://webda.physics.muni.cz}}, and nearby galaxies that are resolved into individual stars \citep{2004yCat.7239....0H, 1973ugcg.book.....N, 1985IAUS..113..541W, 2013A&A...558A..53K}. 
The resulting footprint has an area of ${\sim}\,$18,300\,deg$^2$ and is shown in Figure~\ref{fig:mask}.

We used the galaxy--halo model from \citet{Nadler:2020} to predict the number of satellites within the LSST WFD footprint that would be detectable given our observational selection functions. 
The galaxy--halo model from \citet{Nadler:2020} has eight parameters that associate luminous satellite galaxies with dark matter subhalo properties and model the effects of baryonic physics on subhalo disruption. We summarize the relevant features of that model here.

The galaxy--halo model includes an abundance-matching procedure that monotonically relates the mean satellite galaxy luminosity to subhalo peak maximum circular velocity. 
The model assigns satellite luminosities by abundance matching to the GAMA survey \citep{2015MNRAS.451.1540L} at the bright end ($M_V < -13$\,mag) and performing a power-law extrapolation for fainter satellites with a faint-end slope and log-normal scatter fit to the Milky Way satellite population.
This simple, empirical prescription is not sensitive to the post-infall stellar mass growth of satellite galaxies, but it has been found to be consistent with current data \citep[e.g.,][]{Nadler:2020}.
Furthermore, the galaxy--halo model follows \citet{Graus:2019} in modeling the galaxy occupation fraction (i.e., the fraction of subhalos that host galaxies of any mass) as a sigmoid function defined by a peak mass at which 50\% of subhalos host galaxies and a free slope.

Satellite galaxy sizes are generated by extrapolating a modified version of the mean galaxy size to halo virial radius relation of \citet[][]{2013ApJ...764L..31K} to fainter magnitudes. This model is implemented as a power-law relationship parameterized by an amplitude and power-law index with a scatter implemented as an additional free parameter \citep{Nadler:2020}.  The size model corresponds to satellite galaxy sizes before accretion onto the Milky Way, since post-infall size evolution due to tidal stripping is found to have no appreciable impact on the observed satellite size distributions even in the most extreme cases (e.g., Appendix A.4 of \citealt{Nadler:2020}; \citealt{Errani:2020}; \citealt{Kim:2021}).

Finally, the model accounts for subhalo disruption due to the tidal influence of the Galactic disk. This effect is captured as the probability of disruption, modeled by applying a random forest algorithm trained on hydrodynamic simulations of subhalo populations \citep[][]{2017MNRAS.471.1709G, 2018ApJ...859..129N}. We note that this model may need to be revised in the future given the complex relationship between the disruption of subhalos and the observed stellar component of the galaxies that they host \citep[e.g.,][]{Shipp:2024}.

To predict the characteristics of the Milky Way satellite galaxy population that will be observable with LSST, we sample from the posteriors of the galaxy--halo model parameters that were derived by fitting to data from DES and PS1 \citep{Nadler:2020}. 
This analysis compares the observed data to model predictions from two Milky-Way-like host halos simulated using high-resolution dark matter-only zoom-in simulations \citep[][]{Mao_2015}. The Milky Way satellite population deviates from what is expected from halos of similar mass due to the presence of the LMC, which contributes its own satellite population to the Milky Way population \citep[][]{1976MNRAS.174..695L, 2008ApJ...686L..61D, 2016ApJ...830...59L, 2017MNRAS.472.1060D}. Thus, the two host halos were selected to have an LMC analog with realistic internal and orbital properties. The host halos have virial masses of $1.26$ and $1.57 \times 10^{12} M_{\odot}$, which does not fully cover the plausible mass range of the Milky Way $[1.0, 1.8] \times 10^{12} M_{\odot}$ \citep[][]{2019MNRAS.484.5453C, Cautun_2020, 2019ApJ...886...69L}. Since the subhalo abundance scales linearly with the Milky Way mass, the model is unable to capture the full possible extent of the Milky Way satellite population. However, recent efforts to simulate a larger family of tailored Milky-Way-like simulation (including the full range of likely Milky Way halo mass) should allow more accurate probes of the subhalo abundance at small scales in future work \citep[e.g.,][]{Buch_2024}.

\begin{figure*} 
\centering
    \includegraphics[]{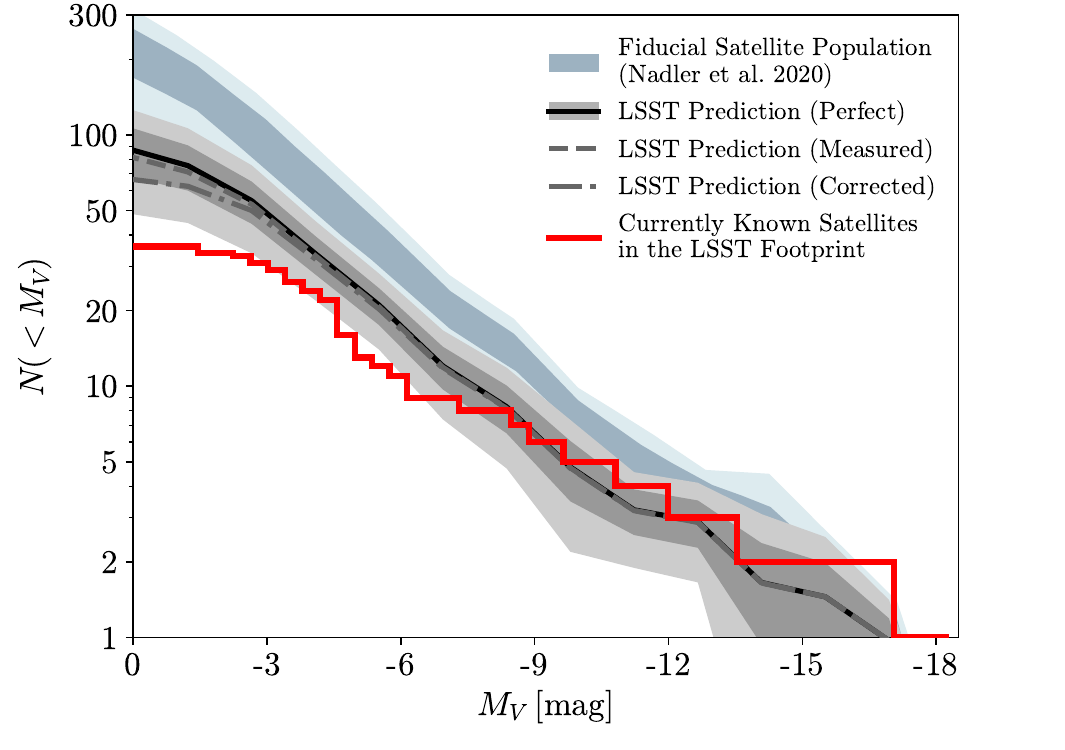}
    \caption{\label{fig:lum_func} Predicted luminosity function of Milky Way satellite galaxies  with $M_V < 0$\,mag and $r_{1/2} > 10$\,pc located within $300\, {\rm kpc}$ (cumulative number of satellite galaxies brighter than a specific $V$-band absolute magnitude). The blue contours show the $1\sigma$ and $2\sigma$ uncertainty on the total Milky Way satellite luminosity function inferred by \citet{Nadler:2020}.
    The black line and gray contours show the mean, $1\sigma$, and $2\sigma$ uncertainty on the predicted luminosity function of satellite galaxies that are detectable by LSST assuming perfect star/galaxy classification. The gray dashed (dot dashed) lines show the mean expectation assuming star/galaxy classification efficiency measured (corrected) in the DC2 simulations. The red solid line shows the luminosity function of currently known satellite galaxies within the LSST WFD footprint as collected in \citet{Pace:2024}.}
\end{figure*}

We present the predicted LSST luminosity function in Figure~\ref{fig:lum_func}, which shows the cumulative number of satellites brighter than a specific absolute magnitude. We show the population statistics with different weights applied based on the detection probabilities for each of the scenarios that our machine-learning classifier is trained on: using perfect star/galaxy separation, observed star/galaxy separation and a corrected observed star/galaxy separation.
We sample the posteriors on the galaxy--halo model parameters $10^4$ times, with each sample returning two Milky-Way-like satellite population realizations.
Since our galaxy--halo model is conditioned on the existence of the LMC, the predicted luminosity function is restricted to always detect the LMC, hence errors are not Poissonian at the bright end. The galaxy--halo model from \citet{Nadler:2020} predicts that the Milky Way has $220 \pm 50$  satellite galaxies with $M_V \lesssim 0$\,mag and $r_{1/2} > 10$\,pc within $300$\,kpc. 
We apply the LSST WFD footprint and survey selection function to predict the number of satellites that will be detected by LSST.
We find that LSST will detect \ntotideal satellite galaxies assuming perfect star/galaxy separation or \ntotmeas (\ntotcorr) satellite galaxies assuming measured (corrected) star/galaxy separation, where ``corrected'' refers to the higher satellite detection significance required to reduce the false positive rate (Section~\ref{sec:results}). These values correspond to the mean and standard deviation calculated by sampling from the posterior of the \citet{Nadler:2020} model and applying our observational selection function.  Given that \nobs satellite galaxies are already known to exist within the LSST WFD footprint \citep{Pace:2024}, this amounts to the predicted discovery of \nideal, \nmeas, or \ncorr satellite galaxies by LSST assuming perfect, measured, or corrected star--galaxy separation, respectively.
To put these numbers in context, the LSST \texttt{baseline} v4.0 WFD footprint covers ${\sim}\,44\%$ of the sky, and we find that with perfect star--galaxy separation, LSST would be able to detect ${\sim}\,90\%$ of the Milky Way satellite galaxies with $M_V \lesssim 0$\,mag, $r_{1/2} > 10$\,pc, and $D < 300$\,kpc in this footprint.

\begin{figure*} 
\centering
    \includegraphics[]{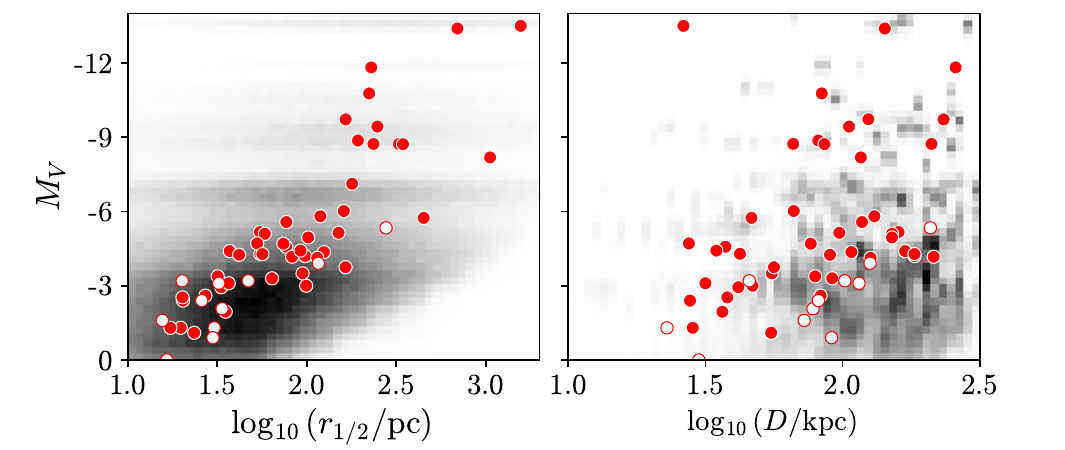}
    \caption{\label{fig:dwarfs} Model predictions for the properties of Milky Way satellite dwarf galaxies detectable with LSST (normalized gray scale). Overplotted are the currently known candidate (open circles) and confirmed (filled circles) Milky Way satellite dwarf galaxies \citep[][and references therein]{Pace:2024}. The left panel shows absolute $V$-band magnitude as a function of size, while the right panel shows absolute $V$-band magnitude as a function of distance. Discrete structure in the simulated satellite population come from re-sampling two simulations of Milky Way-mass host halos possessing LMC analogs (Section~\ref{sec:lum_func}). The model predicted distribution suggests new LSST detections should feature systems with lower luminosities, fainter surface brightnesses and larger distances.}
\end{figure*}

We further present predictions of the luminosities, sizes, and heliocentric distances of the satellite galaxies detectable by LSST in Figure~\ref{fig:dwarfs}.  Each simulated satellite galaxy is weighted by its probability of detection in an idealized scenario with perfect star/galaxy separation. Over-plotted are the currently known confirmed and candidate satellite dwarf galaxies. These results suggest that we should expect LSST to detect satellite galaxies that are fainter and farther away than the currently known satellite population. This will allow us to better study the threshold of galaxy formation at distances where satellite galaxies are less affected by the tidal influence of the Milky Way disk.

We note that the quantitative predictions performed in this section have been limited to the population of Milky Way satellite galaxies. Our observational selection functions could be used to make similar predictions for the population of Milky Way outer halo star clusters; however, making such predictions would require a model of the underlying population of these systems.

\section{Conclusions}
\label{sec:conclusions}

We present projections for the sensitivity of the Rubin Observatory LSST to resolved ultra-faint satellite galaxies and ultra-faint star clusters in the outer halo of the Milky Way. 
We use a combination of the LSST DESC end-to-end simulated data, DC2, and a dedicated suite of catalog-level simulated stellar systems. Assuming perfect star/galaxy separation, we find that a conventional isochrone-based matched-filter search on the LSST data will have $>50\%$ detection efficiency for faint ($M_V \sim 0$\,mag), compact ($r \sim 10$\,pc) satellites out to a distance of $\sim 250$\,kpc. However, we also demonstrate that there is noticeable room for improvement when it comes to optimizing the efficiency of star/galaxy separation. Using a simple star/galaxy classification based on measured quantities, we find that the false positive rate is around 17\% at $D<128$\,kpc and this increases appreciably to 29\% at farther distances. Assuming the best-fit galaxy--halo model and numerical simulations described in \citet{Nadler:2020}, we predict that the 10-year LSST WFD survey will contain \ntotideal, \ntotmeas, or \ntotcorr detectable satellite galaxies assuming perfect, measured, or corrected star/galaxy separation, respectively. LSST will preferentially discover satellites at larger distances, lower luminosities, and fainter surface brightnesses. This new population of satellites will help advance our understanding of the threshold of galaxy formation.

While DC2 is the highest-fidelity simulation of LSST to date, there are several limitations in using it to derive the LSST survey selection function for Milky Way satellites. 
First, DC2 only covers a small region of the high-Galactic-latitude sky and does not capture the full range of inhomogeneities expected in the LSST WFD footprint (e.g., stellar density, interstellar extinction, and survey depth). 
To mitigate the impact on our predictions for the observable Milky Way satellite galaxy population, we explicitly mask regions of the sky where the LSST WFD performance is expected to differ significantly from DC2.
Secondly, DC2 was simulated using an out-of-date model for the 5-year LSST survey cadence and environmental conditions. 
The median $g$- and $r$-band depths from DC2 are compatible with more recent simulations of the baseline LSST WFD strategy, and we restrict our predictions to the region of the LSST WFD footprint that matches the DC2 depth.
Finally, our results come from injecting simulated stellar populations of Milky Way satellites into the DC2 catalogs. Recent studies have shown that catalog-level simulations give slightly higher sensitivity than full image-level simulation \citep[e.g.,][]{Zhang:2025}; however, these differences are relatively small in the ranges of satellite size, luminosity, and distance that we study here.

There is significant room to improve star/galaxy classification, both with LSST data alone and in combination with current and future space missions, such as {\it Euclid} \citep{Euclid:2025} and the {\it Nancy Grace Roman Space Telescope} \citep{Spergel:2015}. The Euclid Wide Survey will cover $\sim$\,15,000\,deg$^2$ in a single broad visible bandpass with an angular resolution of ${\rm FWHM} \sim 0.18\arcsec$ and a 5$\sigma$ point-source limiting magnitude of 26.2 AB mag \citep{EuclidWFS,EuclidVIS}. This data product will aid Rubin star/galaxy separation when used independently or through joint-pixel level analyses \citep[][]{Guy:2022}. Furthermore, proposed synergies between Rubin and Roman suggest that Roman could efficiently distinguish point sources from extended sources over the entire sky down to 25 AB mag using the high-throughput near-infrared F146 filter \citep[][]{Han:2023}. 

There also exist novel ideas for improving star/galaxy separation efficiency using multi-wavelength observations and advanced image processing techniques \citep[e.g.,][]{Kov_cs_2015,Sevilla-Noarbe:2018,Muyskens_2022,Bechtol:2025}. Satellite detection efficiency can also be improved by utilizing likelihood-based search algorithms \citep[e.g.,][]{Bechtol:2015, Drlica-Wagner:2020}. These algorithms construct a likelihood function from the product of Poisson probabilities to detect individual stars based upon their spatial positions, measured fluxes, photometric uncertainties, and the local imaging depth, given a model that includes a putative dwarf galaxy and empirical estimation of the local stellar field population.  The combination of the simple matched-filter algorithm applied here with a likelihood-based search approach has been found to greatly reduce false positives \citep{Drlica-Wagner:2020}.

The wealth of information arising from our continued detection of small-scale structure holds valuable insights into galaxy formation, reionization, the formation of heavy elements and dark matter microphysics. 
The discovery and measurement of these systems will continue to be an invaluable probe into the fundamental nature of our universe.

{\noindent{\it Software:} 
Astropy \citep{2013A&A...558A..33A,2018AJ....156..123A},
HealPy \citep{Zonca2019}
healsparse \citep{Rykoff:2019},
Matplotlib \citep{Hunter:2007},
Numpy \citep{harris2020array},
Pandas \citep{reback2020pandas},
Phalanx\footnote{https://phalanx.lsst.io/},
SciPy \citep{2020SciPy-NMeth},
simple \citep{Bechtol:2015},
ugali \citep{Bechtol:2015, Drlica-Wagner:2020}
}

\section*{Acknowledgments}

This paper has undergone internal review in the LSST Dark Energy Science Collaboration.

Author contributions: KT, SM, ADW led the analysis and paper writing.  JLC, and PSF contributed to the analysis and reviewed the paper. KB, EON, and AHGP contributed algorithmic and modeling infrastructure and reviewed the paper. YYM contributed to the creation of the DC2 dataset and truth-match catalog. AJT contributed to the development and support of the RSP.

This research received support from the National Science Foundation (NSF) under grant No.\ NSF DGE-1656518 through the NSF Graduate Research Fellowship received by SM. This research was partially supported by NSF AST-2307126 and AST-2407526. This research was supported in part by grant NSF PHY-2309135 to the Kavli Institute for Theoretical Physics (KITP).

The DESC acknowledges ongoing support from the Institut National de 
Physique Nucl\'eaire et de Physique des Particules in France; the 
Science \& Technology Facilities Council in the United Kingdom; and the
Department of Energy and the LSST Discovery Alliance
in the United States.  DESC uses resources of the IN2P3 
Computing Center (CC-IN2P3--Lyon/Villeurbanne - France) funded by the 
Centre National de la Recherche Scientifique; the National Energy 
Research Scientific Computing Center, a DOE Office of Science User 
Facility supported by the Office of Science of the U.S.\ Department of
Energy under Contract No.\ DE-AC02-05CH11231; STFC DiRAC HPC Facilities, 
funded by UK BEIS National E-infrastructure capital grants; and the UK 
particle physics grid, supported by the GridPP Collaboration.  This 
work was performed in part under DOE Contract DE-AC02-76SF00515.

NSF–DOE Vera C.\ Rubin Observatory is a Federal project jointly funded by the NSF and the Department of Energy (DOE) Office of Science (SC), with early construction funding received from private donations through the LSST Corporation. The NSF-funded LSST (now Rubin Observatory) Project Office for construction was established as an operating center under the management of the Association of Universities for Research in Astronomy (AURA). The DOE-funded effort to build the Rubin Observatory LSST Camera (LSSTCam) is managed by SLAC National Accelerator Laboratory (SLAC).

This material or work is supported in part by the National Science Foundation through Cooperative Agreement AST-1258333 and Cooperative Support Agreement AST1836783 managed by the Association of Universities for Research in Astronomy (AURA), and the Department of Energy under Contract No.\ DE-AC02-76SF00515 with the SLAC National Accelerator Laboratory managed by Stanford University.

This research has made use of the WEBDA database, operated at the Department of Theoretical Physics and Astrophysics of the Masaryk University.
This research made use of arXiv.org (\url{https://arXiv.org}) and NASA's Astrophysics Data System for bibliographic information.

This manuscript has been authored by Fermi Forward Discovery Group, LLC under Contract No.\ 89243024CSC000002 with the U.S. Department of Energy, Office of Science, Office of High Energy Physics.

\bibliography{main}{}
\bibliographystyle{aasjournal}

\appendix

\section{Rubin Science Platform Performance}

\label{app:rsp}

Our analysis of the DC2 data is performed on the Rubin IDF using the RSP \citep{RSP_ADASS2021}, an online service that enables users to access and analyze Rubin LSST data through a collection of inter-connected Aspects. The entirety of this analysis was conducted from within the Notebook Aspect, which provides access to Jupyter Notebooks and allows for nearly seamless sharing of code snippets and reproducibility of results between collaborators. In particular, we worked with catalog data retrieved via TAP queries to the Qserv database. We note that the Notebook Aspect also provides access to the full LSST Science Pipelines software stack. With the LSST Science Pipelines, users can access datasets through the Butler and run pipelines and tasks to process data. 

The IDF is a shared resource among the LSST user community, which places limits on the available computational power and memory available to any one user. At the time of this analysis, the maximum resource allocation was 32 CPUs and 16GB RAM per user. 
As a result, when running our analysis on the IDF some steps needed to be taken to ensure reasonable turnaround times between iterations of the analysis, while also freeing up sufficient resources for other users to access. Some choices are inconsequential when trying to maintain realism in the analysis: e.g., limiting the size of the queried DC2 region, injecting satellites one-by-one into the DC2 field and running the search. Since \simple operates with a limited search aperture of 0.5 deg and can only scan one region at a time, these choices have no impact on the final results of the analysis. However by limiting the size of our queried region we can increase the speed of each query and avoid maxing out the available memory. Similarly by loading our satellites in one at a time we alleviate the burden on the system memory. 

One choice is more notable: supplying \simple with the true distance moduli and locations of the satellites during the search as opposed to requiring \simple to scan over distance moduli. When applied blindly to search real data, we would perform the search over a range of different distance moduli \citep[e.g.,][]{Bechtol:2015}; however, doing so would have resulted in a runtime of  weeks to months for the $10^5$ simulated satellites. Running \simple at a particular location and at a particular distance modulus takes on average 1.3 seconds. When allowing \simple to cycle through distance moduli, the time increases to about 21 seconds at a particular location, approximately 16 times longer than when a fixed distance modulus is used. Querying times for the RSP TAP server have large variations and depend on several factors such as the number of objects at a region, the server load at a particular time, and unexpected disconnects. Despite the variation in query times, querying the RSP TAP takes up the majority of the time spent during the analysis, leading to a total run time of around two weeks for $10^5$ satellites when providing the true distance modulus for \simple (average of $\sim$12 seconds per object). For a blind search of the Rubin LSST data, we would want to run the full distance modulus scan over the $\sim$20,000\,deg$^2$ WFD footprint. Assuming the same search radius of 2\,deg, the sky could be broken into $\sim$1,600 regions and searching through the data could be completed in less than a day.
We note that these performance characteristics reflect the state of the IDF at the time of our analysis and may not represent future performance as the system evolves and additional resources are made available to users (i.e., through the US Data Facility).

\end{document}